# Strategic Control of Facial Expressions by the Fed Chair[*]

Hunter Ng
Baruch College, City University of New York
hunterboonhian.ng@baruch.cuny.edu

22nd October 2024

**Abstract**

This paper investigates whether the Federal Reserve Chair strategically controls facial expressions during FOMC press conferences and how these nonverbal cues affect financial markets. I use facial recognition technology on videos of press conferences from April 2011 to December 2020 to quantify changes in the Chair's nonverbal signals. Results show that facial expressions serve as a separate public signal, distinct from verbal content. Using deepfakes, I find that the same facial expressions expressed by different Fed Chairs are interpreted differentially. As their tenure increases, negative expressions become more frequent, eliciting adverse market reactions. Furthermore, the market's interpretation of these expressions evolves over time, suggesting that investors process facial cues with dual-processing finite-state Markov memory. In line with the Fed's goals of transparency and non-volatility, I find that Fed Chairs do not strategically control their expressions.

[*]I am grateful to Lin Peng, Boragan Aruoba, Sophia Kazinnik, Svenja Dube, Monica Neamtiu for helpful comments. I also received valuable feedback from presentation participants at Baruch College, City University of New York.

# 1. Introduction

Recent studies use facial expression recognition technology and find that nonverbal communication by the Fed Chair is an important investor information source (Curti and Kazinnik, 2023). Since the facial expressions of the Fed Chair has been established to provide a signal to the market (Curti and Kazinnik, 2023), it is thus an empirical question whether these signals are strategic. In this article, I study whether the Fed Chair strategically controls her facial expressions during FOMC press conferences.

In the modern era of central bank transparency, effective communication by the Federal Reserve is crucial for shaping market expectations and delivering information to markets, especially during economic downturns (Woodford, 2001; Yellen, 2016). The introduction of post-FOMC press conferences in 2011 marked a significant development in how the Federal Reserve communicates, providing a platform for clarifying policy decisions and conveying the underlying motivations behind them. The primary purpose of FOMC press conferences is to communicate the Fed's policy intentions transparently and to reduce market uncertainty (Bernanke, 2013). Under certain conditions, increased transparency is shown to be beneficial, as it aligns market expectations with the central bank's dual mandate, enhancing welfare (Geraats, 2009). To reinforce policy goals, it is in the Fed Chair's interest to convey inflation projections and monetary policy guidance with clarity and a gradualist approach (Stein, 2014).

While the verbal content of FOMC statements has a significant impact on market prices, content related to policy decisions tend to move prices the most, reflecting a strong informational effect (Gomez-Cram and Grotteria, 2021). However, apart from linguistics, facial expressions offer another fundamental source of information (Darwin, 1872). Advances in facial recognition technology have expanded research into non-verbal communication cues, with studies by Curti and Kazinnik (2023) and Alexopoulos et al. (2024) documenting a correlation between market performance and the Fed Chair's facial expressions during events like FOMC press conferences and congressional testimonies. If facial expressions influence market reactions, a key question arises: are these expressions strategically used by the Fed Chair to achieve policy objectives?

I use FOMC press conference videos from 27th April 2011 to 16th December 2020, with facial recognition technology and machine learning algorithms to quantify changes in the Chair's delivery and market responses. My results show that facial expressions serve as a separate signal apart from verbal content. Next, using Deepface, an off-the-shelf facial expression recognition software (Kaur et al., 2022), and training a deepfake model that only replaces face but not emotions, I find that the emotions registered by the software differs significantly, corroborating the investor difference-in-opinion model.

Additionally, I find that investors react negatively to the Fed Chair's negative expressions while controlling for the content in their words. Next, my main result is that from the perspective of negative expressions causing market volatility, the Fed Chairs do not strategically control their expressions. As Fed Chairs increase in their tenure and experience, the frequency of negative expressions increase, which would not be intuitive if there were strategic control since negative expressions cause adverse market reactions. However, there is evidence of market learning from the Fed Chair's expressions as the investors' reactions decrease over tenure time. The results are robust to the Fed Chair's age and other control variables.

Lastly, I find evidence that when a congressional testimony is held just before the FOMC press conferences, the market's interpretation of the negative facial expressions becomes more negative. I also use two of the seven basic emotions identified with Deepface to construct a transparency measure. These results show three findings - (1) that the Fed Chair does not strategically control facial expressions, (2) that facial expressions serve as a separate public signal from words, and (3) that there is investor learning with an optimized finite-state memory.

This article is related to three strands of literature. Firstly, there is a growing economics empirical literature examining information signals of the Federal Reserve using sophisticated technologies and high-frequency data (Gomez-Cram and Grotteria, 2021; Curti and Kazinnik, 2023; Alexopoulous et al., 2024; Swanson and Jayawickrema, 2023; Gorodnichenko et al., 2023). These papers use novel ways of converting non-numerical information such as textual, nonverbal or other types of data in the Federal Reserve's arsenal to examine their effects on information transmission, which is revolutionizing the way we understand how investors process macroeconomic information.

Next, there is a stream of financial economics literature on how investors incorporate signals. Traditionally, central banks have used full-information rational expectations models to guide monetary policy in the wake of the Lucas's (1972) imperfect information model (Calvo, 1983; Coibion, Gorodnichenko and Kamdar, 2018). Gomez-Cram and Grotteria (2022) conduct a thorough comparison of eight frameworks, including Fed put (Cieslak and Vissing-Jorgensen, 2020) and learning about parameter uncertainty. They find that instead of the full-information rational expectation model, Allen, Morris and Shin's (2006) difference-in-opinion model best explains the interpretation of the public signal of FOMC's words (Banerjee et al., 2009). The difference-in-opinion model shows that investors agree to disagree on prices, which can be either common knowledge or uncertainty leading to higher-order beliefs. New models also show how algorithmic analysis has changed disclosure content, which changes the final signal that investors receive and process (Cao et al., 2023).

Lastly, this article relates to use of deepfakes in interdisciplinary research. Westerlund (2019) provides a thorough review of how deepfakes can be used in society. The hyper-realistic videos are created using AI and can digitally recreate actual people, for example, the Fed Chair giving a speech. Emett et al., 2024 find that investors react to deepfake financial news using a realism heuristic and struggle with analytical judgements. Rienert et al., 2024 show that deepfakes can be used to recreate nonverbal behavioral studies.

In comparison with the previous literature, this article makes three contributions to the literature. Firstly, this article finds that following the FOMC committee's goal of transparency and non-volatility, then Fed Chairs do not strategically control their facial expressions. This could provide new insights into monetary policy, as facial expressions could be tapped on as a new policy tool. This is even more important given that the Federal Reserve is constantly looking for new levers in its policy options (Bernanke, 2007).

Secondly, this article finds that complex expressions can have significant market activity effect. Existing literature focuses on negative expressions and its correlation to negative market movement. However, other expression combinations such as transparency can affect markets. As facial emotion recognition software improves in recognizing more types of emotions, there is potential to investigate more complex combinations.

Thirdly, this paper shows that the Fed Chair's facial expressions may be interpreted by investors using a dual-processing, Markov model (Wilson, 2014). In other words, the market learns from the facial expressions and react less as time goes back. This reaction is recalibrated whenever there contemporaneous events, such as congressional testimonies. This is different from classic macroeconomic papers which assume Bayesian rationality with infinite memory. Instead, the bounded Markov model explains the results in this article better. Additionally, this article also showcases the preliminary use of deepfakes in economics to reconstruct and test nonverbal cues.

The remainder of this paper is structured as follows: Section 2 reviews relevant literature on central bank communication and nonverbal cues, and develops the hypotheses. Section 3 describes the data and methodology. Section 4 presents the empirical results, and Section 5 concludes with a summary of my key insights, caveats and suggestions for policy-making.

## 2 Hypothesis Development

**2.1. FOMC Press Conference and Non-verbal Communication**

The FOMC Committee holds eight meetings a year to discuss monetary policy actions. Since May 1999, the FOMC Committee started issuing post-FOMC meeting statements which specified target levels for the federal funds rate. After the federal funds rate hit 0% in December 2008, former Chair Ben Bernanke decided in 2011 to give press conferences after select meetings as an additional policy tool, and since 2019, every FOMC meeting has been followed by a press conference. These post-FOMC press conferences are the focus of this study.

The market responds to post-FOMC press conferences. Lucca and Moench (2015) show a pre-FOMC meeting stock price drift while Boguth et al. (2019) show that this price drift occurs only when a post-FOMC press conference is held. Amengual and Xiu (2018) find that market volatility decreases after the announcement. The overwhelming evidence shows that post-FOMC press conference conveys information to the markets.

This reaction is not only limited to before or after the post-FOMC press conferences. The market responds concurrently during the press conference. Gomez Cram and Grotteria (2022) timestamp the words pronounced in the press conference and align them with high-frequency financial data, showing a positive correlation between changes in the newly-issued policy statement and stock returns.

Facial expressions are a primary way to express emotional states. From psychology, in his seminal work on nonverbal communication, Mehrabian (1972) puts forth the 7-38-55 rule, where 7% of a message is conveyed through words, 38% through vocal elements, and 55% through nonverbal elements such as facial expressions. Zhang et al. (2023) presents Emotion-CLIP, a vision-language pre-training paradigm that uses visual and verbal elements to improve emotion recognition. They show that understanding emotions involves integrating both verbal and nonverbal cues.

Existing literature in finance and accounting builds on this and finds correlation between nonverbal cues and investor reactions[1]. Mayew and Venkatachalam (2012) show that a manager's voice in earnings conference calls conveys important information to investors. They provide evidence that a stressed voice indicator is a better indicator of future firm performance than the words in the conference call. Davila and Guasch (2021) interestingly study nonverbal behavior in entrepreneurs pitching their business ideas and measure body expansiveness using OpenPose[2]. An individual has an expansive posture if they have widespread limbs, a stretched torso, and/or take up considerable physical space. An expansive posture is associated with higher forecast errors, higher likelihood of funding success, but a lower survival rate. The authors argue that expansiveness is correlated with dominance and attractiveness, which affect investor reactions.

Economics research has also tapped on rich facial expressions datasets. Breaban and Noussair (2018) found in an experimental setting that traders' facial expressions of fear are linked to negative stock price movements, while positive emotions are associated with overpricing. This suggests that experienced traders can exhibit nonverbal cues that may mislead market participants. Alexopoulous et al. (2023) show that the Chair's facial expressions during congressional testimonies have significant effects on the markets, reaching magnitudes comparable to those after a policy rate cut. The evidence points to a significant effect of facial expressions on markets.

---

[1] Refer to Hanlon et al., 2022 for a comprehensive review of the literature.

[2] This is an algorithm developed by the CMU Perceptual Computing Lab. It uses a Unity plugin that jointly detects human body, hand, facial and foot keypoints on single images.

Curti and Kazinnik (2023) use facial recognition software to analyze the change in facial expressions of the Chair. They find that the negative facial expressions are correlated with negative stock returns after controlling for negative tone in the words, showing that the market infers additional information from the Chair's non-verbal cues. Taken together, the evidence points to the fact that facial cues for each Fed Chair may be a complex signal. In other words, if two Fed Chairs try to emote happy expressions, they could be perceived differently by the market.

Following this, I state my first hypothesis in alternate form.

**H1: The same facial expression by different Fed Chairs are interpreted differently**

**2.2. Transparency, Non-volatility and Conscious Control**

In psychology, Lambie and Marcel (2002) posit a theoretical framework where there are three levels of emotional awareness. The higher levels of emotional awareness are conceptual and reflective awareness, where individuals can have more sophisticated emotion regulation strategies. Such strategies include cognitive reappraisal and suppression, where we can alter our outward sign of emotion, for example, by hiding negative feelings during a professional setting. This framework helps us to understand that as an individual develops higher levels of emotional awareness and is more experienced, she can consciously control the display of facial expressions. In the case of the Fed Chair, whether or not she consciously controls facial expressions in order to achieve the Fed's goal has not been studied.

Hence, to check whether there is conscious control or not, I assume alignment to the Federal Reserve's goals. FOMC press conferences were introduced with the purpose of increasing transparency (Bernanke, 2013). Blinder (1998) explains that by making itself more predictable to the markets, the central bank makes market reactions to monetary policy more predictable. Bernanke (2013) expounds on the changing roles and functions of the Federal Reserve and points out that financial stability is a longstanding goal of the central bank. Amador and Weill (2012) show that in a a large economy with dispersed private information, one way of optimal communication is to be fully transparent. As such, large fluctuations and contemporaneous overreactions to the FOMC press conference are incompatible with the goal of increasing predictability and transparency.

From behavioral sciences, Stouten and De Cremer (2010) show experimentally that happy faces elicit trust and feelings of honesty and transparency. On the other hand, when people's facial expressions communicate anger, they are more likely to be interpreted as less trustworthy and cooperative. This in turn causes their verbal communications to be evaluated as less honest and meaningful. Dotsch and Torodov (2012) and Hsieh et al. (2019) show that facial features and by extension, facial expressions of happiness conveyed trustworthiness and transparency. Given that the FOMC press conference should increase transparency and reduce market volatility, I hypothesize that the Fed Chair is strategic in his/her facial expressions.

**H2: The Fed Chair strategically controls negative facial experience during FOMC press conferences**

**2.3. Processing of nonverbal cues and financial information**

From psychology, the dual processing framework explains how investors process Fed Chairs' facial expressions within a bounded Markov model (Wilson, 2014). This cognitive theory divides mental functioning into two distinct systems: System 1, which is fast, automatic, and emotionally driven, and System 2, which is slower, deliberate, and analytical (Kahneman, 2011). When a Fed Chair makes a significant public appearance, such as during a FOMC press conference, investors' System 1 is activated by the immediate and emotionally charged nature of the nonverbal cues. This rapid, heuristic-based processing leads to a quick adjustment of beliefs regarding monetary policy, with a strong initial reaction that aligns with the bounded memory assumption. The bounded Markov model reflects this dynamic by allowing the influence of these facial expressions to be high right after the event, but gradually decay as time passes and the emotional salience of the expressions fades in the absence of new supporting information. Over time, as the impact of System 1 diminishes, System 2 processing takes over, providing a more reflective and analytical assessment of the broader economic outlook, which leads to a reduced weight placed on the initial emotional response.

In contrast, traditional Bayesian models with infinite memory imply that investors constantly update their information signals based on the Fed Chair's expressions, regardless of how much time has passed since they started reading the Chair's expressions.

The dual processing framework in a Markov model setting posits that investors may initially react strongly to a Fed Chair's facial expressions during a speech, but as time passes and more economic data or policy information becomes available, these early reactions lose their potency. However, recalibration can occur during significant events that provide new information, such as congress testimonies. In these moments, both emotional and rational processing can be re-engaged, recalibrating investor expectations based on fresh cues. This framework provides a behavioral model of understanding how investors interpret facial cues.

Figure 1 shows a deepfake model, which overlays Chair Powell's face on Chair Yellen's. Using facial expression recognition (FER) software, both register different emotions, even though the created deepfakes are trained to only replace the face but replicate emotions (Li and Deng, 2022). Thus, the same expression could potentially be interpreted as different signals.

**H2: Markets use a dual-processing Markov model with interpreting the Fed Chair's expressions during FOMC press conferences**

## 3. Data

### 3.1. Minute-level Market Data

I first look at economic data of asset classes which are affected by FOMC press conferences. High-frequency changes in asset prices addresses endogeneity issues through the narrow time windows, which decreases the likelihood of other information affecting the asset prices (Cochrane and Piazzesi, 2002; Nakamura and Steinsson, 2018). This approach offers a granular and targeted approach at the underlying factors behind market movement.

I use the following variables to measure the market reaction to the nonverbal cues by the Chair during the FOMC press conference (from January 2011 to December 2020).

- SPDR S&P 500 (SPY): Minute-by-minute SPY prices and SPY trading volumes (number of shares traded).

- CBOE Volatility Index (VIX): Chicago Board Options Exchange Market Volatility Index (VIX), which measures implied volatility of the S&P 500.
- Euro-to-USD Exchange Rate (EUR): Minute-by-minute data for the Euro-to-USD exchange rate and its tick count per minute
- Japanese Yen-to-USD Exchange Rate (JPY): Minute-by-minute data for the JPY-to-USD exchange rate and its tick count per minute

Based on the data, I calculate percent changes within 3-minute intervals in SPY, VIX, EUR, and JPY prices, all measured in basis points. I also calculate average trading volumes within these intervals during press conferences. Table 1 defines these variables, while Table 2 provides descriptive statistics. Table 2, Panel A provides descriptive statistics for the 3-minute interval price changes.

### 3.2. Interpreting Nonverbal Cues

State-of-the-art computer vision and machine learning lets us automatically recognize and quantify facial expressions with high accuracy and scalability. They allow for standardized yet dynamic measurements compared to a human. To measure the Chair's expressions, I use DeepFace, an open-source, lightweight face recognition and facial attribute analysis which uses a five-stage pipeline of "detect, align, normalize, represent and verify" to analyze facial expressions from press conference videos.

DeepFace wraps many state-of-the-art face recognition models - *VGG-Face , FaceNet, OpenFace, DeepFace, DeepID, ArcFace, Dlib, SFace and GhostFaceNet*. I use DeepFace v0.0.91, which is last updated in 2024 and the default configuration *VGG-Face model*. It uses a convolutional neural network (CNN) which represents faces as multi-dimensional vectors. DeepFace returns emotional scores for seven facial emotions (*angry, fear, neutral, sad, disgust, happy and surprise*), where each emotion is scored from 0 to 100 and they sum up to 100. Fig. 2 provides an example of the measured frame.

To prepare the video data, I first download the FOMC press conferences from the Federal Reserve's Youtube Channel. I then convert each video into a set of frames with two-second intervals using a python Flask server. To measure the facial expressions, I use *DeepFace v0.0.91*, which identifies the facial expressions and scores them. Once the frames are scored, I aggregate them into a rolling three-minute level, following the approach of (Curti and Kazinnik, 2023). I then take the average score of a particular emotion, e.g. happy, in a three minute interval, to get a measure of the happiness expressed by the Chair during the interval. I also get the total lifetime average score across all FOMC press conferences of each emotion by the different Fed Chairs. This method creates a baseline for the emotions because each Fed Chair's facial structure affects how the facial software perceives it. For instance, if Chair Powell's face is predisposed to be angry all the time, this method ensures a fair comparison among the three.

In particular, to measure the Negative and Transparent Facial Expressions, I use the following specification.

$$\text{Negative Facial} = \frac{\text{Angry}_{\text{3-mins average}} + \text{Fear}_{\text{3-mins average}} + \text{Disgust}_{\text{3-mins average}}}{\text{Angry}_{\text{chair lifetime average}} + \text{Fear}_{\text{chair lifetime average}} + \text{Disgust}_{\text{chair lifetime average}}}$$

$$\text{Transparent Facial} = \frac{\text{Happy}_{\text{3-mins average}} + \text{Neutral}_{\text{3-mins average}}}{\text{Happy}_{\text{chair lifetime average}} + \text{Neutral}_{\text{chair lifetime average}}}$$

My initial sample includes 2657 minute-level observations from 46 FOMC press conferences held between April 2011 and September 2020. Of these, there 18 with introductory statements. On average, each press conference lasts about 55 minutes, with the initial 10 minutes typically dedicated to the introductory statement.

Not all the screenshots are of the Fed Chair. To overcome this, I use a pre-trained VGG16 model from the Keras library. VGG16 is a 16-layer CNN model and the weights for the VGG16 model provided by Keras are pre-trained on the ImageNet dataset, which contains millions of images across a thousand categories. I first locate a reference picture on the FOMC press conference of the Chair talking, manually check that this frame is in line with the majority of the frames of the Fed Chair, then use a cosine similarity test with a threshold of 50% to accurately sort the screenshots into whether they are of the Fed Chair or not. Other images could include a reporter asking questions or of a diagram being shown to explain FOMC policies.

To ensure robustness, I only take screenshots where both the VGG16 cosine similarity test and the facial analysis displayed a N.A result. Both programs check for whether the Fed Chair is in the image or whether a face is on screen respectively. This ensures that I am correctly capturing the facial expression of the Fed Chair and not a random screenshot.

As the prices of the financial assets are minute-by-minute and the frames are two-second-by-two-second, the optimal way is to aggregate the emotion readings for each frame to the minute of each financial reading. However, there are many frames at the minute-level mark that focus on a reporter or a background image. To ensure robustness and preserve algorithmic integrity, I discard these screenshots and reach a final sample of 1440 minute-level observations.

**3.3. Press Conference Timestamps and NLP Methods**

I now analyze the verbal content of press conferences to accurately assess the impact of facial expressions and control for textual characteristics of the words.

To synchronize the text with the video feed to ensure precise alignment, I use OpenAI Whisper, a speech recognition system developed by OpenAI. It uses a transformer architecture and an encoder-decoder structure in its model. The model is trained on a large dataset and overcomes accents and dialects, understand context, maintains coherence over long audio segments, and is able to give the timestamp accurately. I also remove segments during the Q&A where journalist asks questions based on the timestamps given by the 2-second-interval labelled screenshots based on the VGG16 model.

To quantify the verbal component of the press conference, I use a modified version of BERT model called FinBERT, which is a specially catered to parse financial texts (Huang, Hui and Yi. 2023). FinBERT is encoded on the Financial phrase-Bank dataset, which consists of 4845 english articles that were cateogorized by sentiment class and were annotated by 16 reasearchers with a financial background.

FinBERT itself consists of a tokenization set-up and a model. For my purpose, I use the original FinBERT trained by Huang, Hui and Yi. To measure the other linguistics attributes, I use spaCy, a pre-trained LLM model that is lightweight and optimized for NLP functions. I obtain a list of keywords from past literature. More details can be found in the Appendix A3.

$$\text{NLP}_{k,t} = \frac{\text{NLP}_{k,\,t'}}{\text{NLP}_{k,\,\text{average in a day}}}$$

To construct the NLP measures, I first use the NLP technique on a dialogue-sentence attached to each screenshot. $k$ represents each NLP measure, for example, hawkishness, positivity, etc. $t$ represents each minute, and $t'$ represents each minute before averaging. For instance, to measure if the dialogue-sentence is positive or not, I use FinBERT to classify the sentence's sentiments. Since the financial figures and facial expressions are aggregated to each minute at the smallest level, I calculate a rolling-one-minute window of the count of the NLP and the average of all the rolling-one-minute counts in a day. I then take the count of the NLP measure divided by this average to derive the average score for each minute of FOMC press conference video.

**3.4. Deepfakes to measure nonverbal cues' complexity**

To create a deepfake video, I use the open-source DeepFaceLabs. It uses a combination of multi-task Cascaded CNN, autoencoder architectures, and other algorithms to face-swap. It follows the following steps of face detection and alignment, extraction, preprocessing, model training, face swapping and post-processing to minimize loss functions.

I use Principle Component Analysis (PCA) to analyze the FOMC press conference's emotive intensities, grouped by Fed Chair, and find the centroid. I then take the video with the smallest distance from the centroid. This video represents the FOMC press conference that is most representative in emotions of all the FOMC press conference done by the Fed Chair. Chair Bernanke's is FOMC Press Conference March 2013, Chair Yellen's is FOMC Press Conference December 16, 2015 and Chair Powell's is FOMC Press Conference January 30, 2019.

I then use DeepFaceLab to change the face of each Fed Chair to another Fed Chair. The algorithm preserves the underlying facial expressions, down to the movement of the eyebrow and the mouth. I extract one minutes of uninterrupted video from each one and train the model up to about 100,000 iterations. Even on a moderately fast CPU and GPU, I train for about four hours to obtain ten seconds of coherent, superimposed video for each Fed Chair pair. More details can be found in Appendix A4.

**3.5. Other Control Variables**

To account for other omitted variables such as federal fund rates change and online attention of each meeting, I incorporate additional control variables. Table 1 defines these variables, while Table 2 provides descriptive statistics.

Following Curti and Kazinnik (2023), I include the change in the Federal Funds Rate (FFR) for the current FOMC meeting to control for the actual change in the interest rate. Additionally, I include pre-drift variables to control for the documented pre-drift effect (Lucca and Moench, 2015). The SPY, VIX, and EURUSD pre-drift variables represent the percentage change in the respective asset prices within 30 minutes before the FOMC press conference.

I also use the Monetary Policy Uncertainty (MPU) index (Husted et al., 2020). This index measures public uncertainty about Federal Reserve policy actions and their consequences.

## 4. Results

**4.1 Are the same expressions by different Fed Chairs read differently?**

In my preliminary result, I first ask whether facial expressions are a complex signal. Table 3 shows the min-max normalized changes of analyzing a twenty-second video of each Fed Chair that has their faces changed via the deepfake technique. Each deepfake takes approximately 4 hours to train with a given high-spec GPU stated in Appendix A4. The results indicate that even though the underlying expression is the same, when the face is changed, the facial emotion recognition software registers different emotions. *disgust* changes the most between Yellen's original face and overlaying it with Powell's face. Across the board, we see that *sad*, *surprise* and *neutral* are the most stable while the other emotions change more widely.

This shows that there is a baseline of emotions for each Fed Chair and that expressions are actually not so easy to read from person to person, lending support to the macroeconomic difference-in-opinion model proposed by Allen, Morris and Shin (2006) and Banerjee et al., (2009).

**4.2 Do investors react to negative facial expressions?**

I next examine whether Chairs' emotions are related to the changes in stock and currency market. I use a fixed-effects regression model and estimate the following model:

$$\%\Delta\text{Market}_t = \alpha + \beta_1 \text{Negative Facial}_{t\text{-}1} + \text{controls} + \varepsilon_t$$

For each market reaction $\%\Delta\text{Market}_t$, I calculate the absolute percentage change of each 1-minute interval In untabulated results, I find no meaningful results if I use percentage change without the absolute sign. My results replicate Curti and Kazinnik (2023) as they find that expressions cause reactions from the market. I take this to provide robustness to my data collection and methods. If the Fed Chair makes a negative expression, investors may perceive it as good and bad depending on how they interpret it. For example, when Chair Powell says *"We continue to anticipate that ongoing increases in the target range for the federal funds rate will be appropriate in order to attain a stance of monetary policy that is sufficiently restrictive."*, a negative expression could be interpreted as federal fund rates increasing due to the vibrant business economy, which would be interpreted as bullish for the markets. It could also be interpreted as the Fed's increase have negative impact on borrowing and thus, causing the market to become bearish. Table 4 presents my main results. Columns (1)-(9) examine the different specifications of the model and

To further investigate the relationship between the markets and the Fed Chair's facial expressions, I look at trading volume and tick count. I perform a multivariate regression similar to the main regression in this sub-section, but instead, I change the dependent variables to volume.

Table 5 shows that the volume traded likewise corresponds to the Fed Chair's negative expressions.

**4.3 Does the Fed Chair consciously control her facial expressions?**

If facial expressions indeed affect investors' assessment, then the next relevant question would be whether the Fed Chair consciously controls their facial expressions. If the Fed Chair knew that negative reactions cause instantaneous reactions from the market, they should decrease their use of negative expressions while increasing the frequency of other expressions such as neutral or happy.

Table 6, shows that as the Fed Chair tenure increases, the frequency and intensity of their negative expressions increase. Columns (1) and (2) show that conference count directly increases the negative expressions. From this result, I conjecture that the Fed Chairs do not strategically control facial expressions. Since more negative expressions elicit more negative reactions (Curti and Kazinnik, 2023), and that it also goes against the goal of reducing market volatility, Fed Chairs should reduce such expressions if they were strategic[3]. In fact, this problem was recently highlighted in that excessive Federal Reserve communication leads to much undue volatility (El-Erian, 2023). I bear in mind that FOMC press conferences previously under Bernanke and Yellen were held only on select dates but after 2018, they are held eight times a year. Whether this volatility is a conscious effort of the Fed Chair cannot be explicitly tested but in line with the implicit goal of non-volatility, I conclude that there is no strategic use on the Fed Chair's part.

In columns (4) - (7), I show that as the conference count increases, the Fed Chair also decreases the ratio of neutral expressions. The Fed Chair shows more negative expressions and less neutral expressions, while there is no statistically significant relationship for happy or sad expressions.

**4.4 Do investors react to facial expressions that convey transparency?**

The literature often uses only negative expressions or sentiments to quantify the FOMC press conference. Do neutral or happy facial expressions, which convey transparency, affect markets?

Table 7, Columns (1-9) look at whether other facial expressions affect the markets. There is significance in columns (3),(6),(7) for transparent expressions, showing that investors respond to expressions other than negative ones. From an information economics point of view, when the Fed Chair shows transparent emotions, it could be taken as a signal that there is increased willingness of the Chair's part to convey information credibly to the markets. Thus, investors

---

[3] The results are robust to varying economic conditions. It was pointed out that this result could be because of the age of the Fed Chair causing more negative expressions (Grimmer et al., 2021). However, in this dataset, the maximum tenure is four years for Yellen. It is unlikely that they massively aged in this timespan.

react to it. The effect for single emotions such as happiness, sadness or neutrality do not show as much significance, showing that investors interpret emotions on a higher dimension compared to just the seven basic emotions. This reinforces prior literature that facial expressions drive market expressions and my results show interestingly that complex combinations of facial expressions drive more reactions and volatility.

**4.5 Do investors react to facial expressions that contrast with word sentiment?**

From the above subsections, investors only respond to negative facial expressions. However, what is the underlying reason? Is this effect magnified when there is a contrast between what is being said and what is expressed on the face? Or is it just simply that facial expressions boost the severity of negative sentiments in the Fed Chair's words.

To investigate, I perform a fixed-effect regression of facial expressions that are in contrast to the sentiment of the words expressed at the same instant. Table 8 shows that there is no statistically significant relation between either SPY, VIX or EUR with an interacted term of negative sentiment with negative facial expressions. I take this as evidence that investors are not necessarily correlating words with facial expressions. Rather, the facial expressions of the Fed Chair act as a separate signalling tool that is distinct from the content of the words. This is in line with the nonverbal literature of Mehrabian and Darwin.

**4.6 How do investors interpret negative facial expressions?**

Experience affects how the Fed Chair conducts the FOMC press conference. Curti and Kazinnik (2023) show that as the Fed Chair's tenure increases, their negative expressions are better interpreted by the market. However, how do investors interpret these expressions? Do they learn from the expressions and do recent events affect these interpretations?

Table 9, shows that as the Fed Chair's tenure increases, the interaction term between tenure and negative facial expressions shows significance. The result is consistent across the various markets. I interpret this as corroboration of Curti and Kazinnik's results (2023). However, the coefficients are negative in columns (2) and (6). I take this as evidence that the reactions are interpreted by investors with less volatility. This means that the market becomes more sure of the facial expressions, such that they do not react as much in terms of absolute magnitude.

In further tests, if the Fed Chair has just concluded a congressional testimony hearing before the FOMC press conference (Columns 5 and 8), their facial expressions' negativity have a larger effect on markets. I interpret this as congressional testimonies being high-pressure and causing the Fed Chair to be more uptight in the short-term, thus causing their negative facial expressions to be interpreted with much more volatility than usual. This provides evidence that recent events do affect the Fed Chair's ability to use facial expressions, as well as the market's perception of their facial expressions.

Column (3) and (7) also show that as the Fed Chair nears each progress quartile in their career, their negative expressions are read with less importance. This is evidence that the market indeed changes its expectations of the negative facial expressions as time goes by, leading to an experience effect.

Taken in all, there is a strong support for a bounded memory Markov model, where investors learn from the Fed Chair's expressions and their reactions gradually decrease but there is a bounded memory because recent events such as congressional testimonies affect this reaction. The testimony is a "disruption" which causes a transition to a state with increased reaction, reflecting the market's adjustment to new information (Wilson, 2014).

## 5 Conclusion

In this article, I show that the facial expressions are a complex, distinct signalling tool from word content during FOMC press conference. Even when word sentiment contrasts with the facial expressions, investors do not interpret them accordingly. They do not react uniformly to negative expressions and instead, interpret it as if it were a different signalling tool. This reaction is affected by their experience with the Fed Chair, and also by recent events, which points towards a dual-processing, limited memory Markov model. Mostly important, if the Fed Chair's goal is to reduce volatility, then this article concludes that they are not strategically controlling their facial expressions to influence markets.

There are some caveats here. Firstly, because facial expression recognition (FER) software uses machine-learning, different models and different video resolutions may give different readings of facial expressions. Furthermore, machine-learning models often have in-built variation in them so there may be minimal changes from one run to another. Secondly, I aggregate the emotions into 3-minute rolling averages but there is a wide range of emotions possible within this window. Currently, data related to stock trades are on a minute-by-minute level, thus this is a limitation. Trades placed by the market may also experience a delay as investors take time to process the emotions since it is a complex source of data.

# Tables and Figures

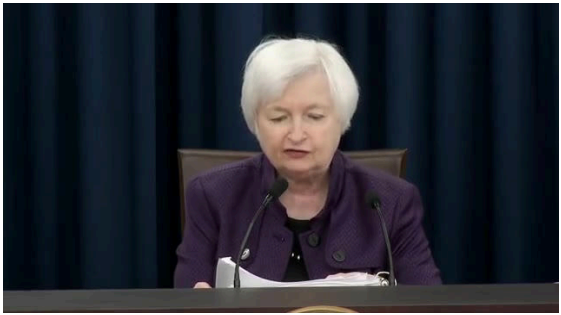

**Panel A:** Fed Chair Yellen speaking during a FOMC press conference. Dominant emotion identified by facial analysis software - **Sad**. Complete Analysis: Angry: 0.722%, Disgust: 0.036%, Fear: 21.992%, Happy: 0.057%, Sad: 58.435%, Surprise: 0.021%, Neutral: 18.737%

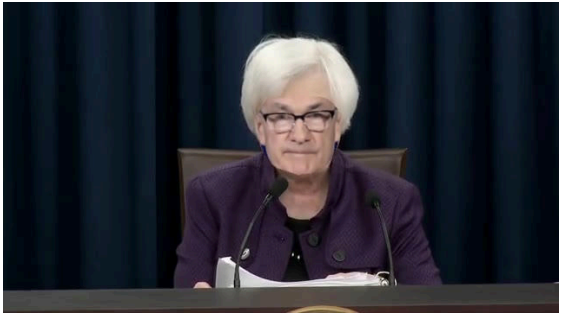

**Panel B:** Deepfake of Fed Chair Yellen speaking, using Fed Chair Powell's face, during a FOMC press conference. Dominant emotion identified by facial analysis software - **Angry**. Complete Analysis: Angry: 99.957%, Disgust: 0.0000038%, Fear: 0.001%, Happy: 0.000%, Sad: 0.041%, Surprise: 0.000%, Neutral: 0.000%

**Figure 1.** Comparison of a deepfake of Fed Chair Janet Yellen during FOMC press conference on September 21, 2016 using Fed Chair Powell and their facial analysis result

| Panel A: Ben Bernanke, April 27 2011 | |
| --- | --- |
| Emotion | Score |
| Angry | 21.0 |
| Disgust | 0.0 |
| Fear | 0.2 |
| Happy | 60.0 |
| Sad | 12.7 |
| Surprise | 0.0 |
| Neutral | 5.8 |

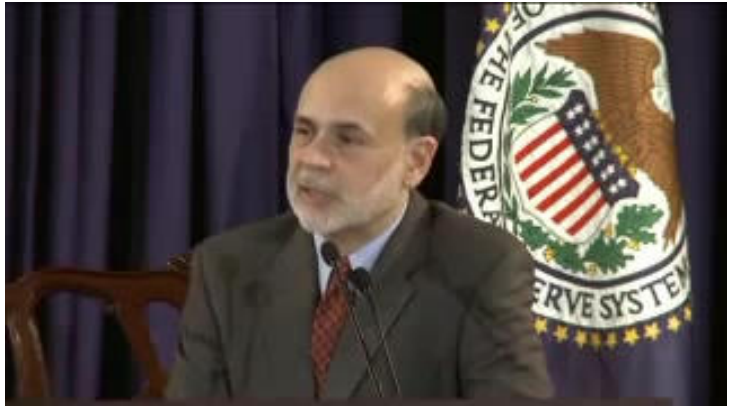

| Panel B: Janet Yellen, June 14, 2017 | |
| --- | --- |
| Emotion | Score |
| Angry | 47.0 |
| Disgust | 0 |
| Fear | 2.6 |
| Happy | 10.2 |
| Sad | 19.3 |
| Surprise | 0.2 |
| Neutral | 20.7 |

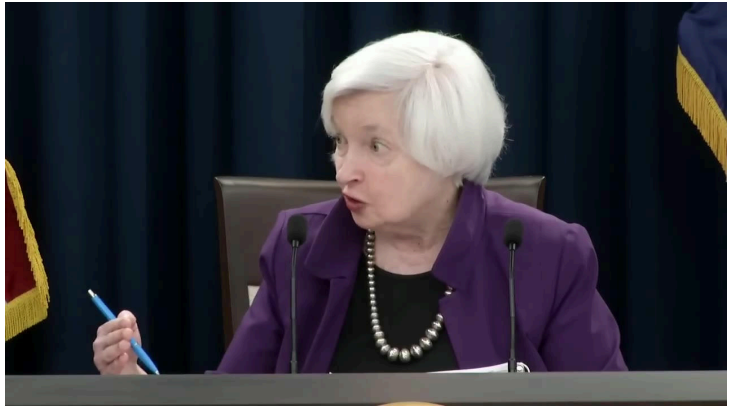

| Panel C: Jerome Powell, Sept 26 2018 | |
| --- | --- |
| Emotion | Score |
| Angry | 0.20 |
| Disgust | 0 |
| Fear | 0 |
| Happy | 0.1 |
| Sad | 2.5 |
| Surprise | 0 |
| Neutral | 97.2 |

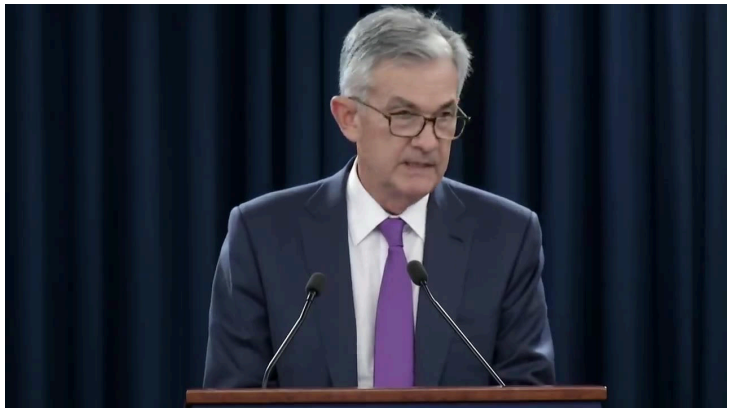

**Figure 2.** Emotion Scores: The emotion intensity scores are captured by DeepFace. Panel A is Ben Bernanke during the FOMC press conference on April 27th, 2011. Panel B is Janet Yellen during the FOMC press conference on June 14th, 2017. Panel C is Jerome Powell during the FOMC press conference on September 26th, 2019.

**Table 1**

Variable Definitions. This table presents definitions of the variables used in the paper.

| Variables | Definition | Source |
| --- | --- | --- |
| %Δ SPY | Percent change in SPY (SPDR S&P 500), measured every minute | Bloomberg |
| %Δ VIX | Percent change in VIX (CBOE Volatility Index), measured every minute | Bloomberg |
| %Δ EUR | Percent change in spot EUR-USD exchange rate, measured every minute | Bloomberg |
| %Δ JPY | Percent change in spot JPY-USD exchange rate, measured every minute | Bloomberg |
| SPY Volume | SPY trading volume in a 1-minute interval | Bloomberg |
| EURUSD Tick Count | EURUSD number of tick counts in a 1-minute interval | Bloomberg |
| JPYUSD Tick Count | EURUSD number of tick counts in a 1-minute interval | Bloomberg |
| Independent Variables | | |
| Negative Facial | Chair's intensity of transparent facial expressions averaged in the prior three minutes divided by average transparent facial expressions across all FOMC meetings by the Chair | DeepFace |
| Transparent Facial | Chair's intensity of transparent facial expressions averaged in the prior three minutes divided by average transparent facial expressions across all FOMC meetings by the Chair | DeepFace |
| Neutral Facial | Chair's intensity of transparent facial expressions averaged in the prior three minutes divided by average transparent facial expressions across all FOMC meetings by the Chair | DeepFace |
| Happy Facial | Chair's intensity of transparent facial expressions averaged in the prior three minutes divided by average transparent facial expressions across all FOMC meetings by the Chair | DeepFace |
| Sad Facial | Chair's intensity of transparent facial expressions averaged in the prior three minutes divided by average transparent facial expressions across all FOMC meetings by the Chair | DeepFace |
| $predrift_k$ | Percent change in $k$ from 2.00pm to 2.30pm, the time between when FOMC statement is released and FOMC press conference is held, where $k$ is SPY, VIX, EURUSD, or JPYUSD. | Bloomberg |

**Table 1**

Variable Definitions. This table presents definitions of the variables used in the paper.

| Variables | Definition | Source |
|---|---|---|
| cfquart | Dummy variable of a chair's FOMC press conference throughout the four quartiles of their count of FOMC press conference | |
| congre30 | Dummy variable of whether a congressional testimony was held within 30 days prior to the FOMC press conference | |
| congre10 | Dummy variable of whether a congressional testimony was held within 10 days prior to the FOMC press conference | |
| %Δ FDFD | Percent change in ICAP US Federal Funds Rate Index on day of FOMC announcement, measured daily | Bloomberg |
| MPU | Monetary Policy Uncertainty (MPU), measured daily | Bloomberg |
| Public_Interest | 3-day average of Google Search Index Value before the actual date of FOMC Press Conference. Recommended Keywords by Google most related to the topic - "FOMC Meeting" | Google |
| NLP Variables | | |
| Negative Sentiment | FinBERT Measure of negativity of a statement. FinBERT is a Large Language Model specialized for financial language. (Huang et al. 2022) | |
| Statement-related | Measures whether the statement is related to the FOMC-statement released at 2.00pm (Gomez-Cram and Grotteria, 2021) I use spaCy, a English language model designed for NLP tasks to tokenize the corpus and search against keywords | |
| Hawkish | Measures in binary format whether statement is hawkish or dovish using word-match (Neuhierl and Weber, 2019) I use spaCy, a English language model designed for NLP tasks to tokenize the corpus and search against keywords | |

**Table 1**

Variable Definitions. This table presents definitions of the variables used in the paper.

| Variables | Definition | Source |
|---|---|---|
| FLS | Measures how forward-looking each statement is based on a list of key words. I use spaCy, a English language model designed for NLP tasks to tokenize the corpus and search against keywords | |

**Table 2**

Descriptive Statistics

| Variable | Count | Mean | Std Dev | Min | 25% | 50% | 75% | Max |
|---|---|---|---|---|---|---|---|---|
| %Δ SPY | 1471 | 0.039 | 0.043 | 0 | 0.012 | 0.028 | 0.050 | 0.538 |
| %Δ VIX | 1471 | 0.272 | 0.391 | 0 | 0.065 | 0.151 | 0.341 | 4.509 |
| %Δ EUR | 1471 | 0.025 | 0.029 | 0 | 0.008 | 0.017 | 0.035 | 0.356 |
| %Δ JPY | 1471 | 0.023 | 0.025 | 0 | 0.009 | 0.018 | 0.031 | 0.259 |
| Volume SPY | 1456 | 0.478 | 0.502 | 0.027 | 0.179 | 0.339 | 0.629 | 11.083 |
| Tick_Count_vix | 1471 | 4.750 | 1.564 | 3.000 | 4.000 | 4.000 | 4.000 | 8.000 |
| EUR Tick | 1471 | 1.919 | 1.427 | 0.247 | 0.982 | 1.480 | 2.379 | 10.378 |
| Negative Facial | 1465 | 1.206 | 0.481 | 0.185 | 0.858 | 1.151 | 1.470 | 3.083 |
| Transparent Facial | 1465 | 1.786 | 0.849 | 0.222 | 1.200 | 1.605 | 2.157 | 4.454 |
| Neutral Facial | 1465 | 1.208 | 0.790 | 0.001 | 0.644 | 1.050 | 1.600 | 5.654 |
| Happy Facial | 1465 | 1.233 | 1.203 | 0.000 | 0.417 | 0.900 | 1.679 | 9.860 |
| Sad Facial | 1465 | 1.194 | 0.766 | 0.004 | 0.648 | 1.058 | 1.577 | 5.385 |
| Predrift SPY | 1456 | 0.126 | 0.408 | −1.020 | −0.163 | 0.093 | 0.366 | 1.060 |
| Predrift VIX | 1426 | −1.958 | 3.431 | −13.009 | −3.226 | −1.220 | 0 | 4.618 |
| Predrift EUR | 1471 | 0.059 | 0.325 | −0.874 | −0.042 | 0.072 | 0.156 | 0.920 |
| Predrift JPY | 1471 | 0.020 | 0.317 | −0.836 | −0.131 | −0.038 | 0.196 | 1.017 |
| Negative Sentiment | 1471 | 0.700 | 0.726 | 0 | 0 | 0.526 | 1.138 | 3.699 |
| Statement_Related | 1410 | 1.134 | 3.729 | 0 | 0 | 0 | 0 | 64.000 |
| %Δ FDFDperChange | 1471 | 0.030 | 0.142 | −0.417 | 0 | 0 | 0.100 | 0.610 |
| Hawkish | 1471 | 0.552 | 5.028 | 0 | 0 | 0 | 0 | 66.500 |
| Public_Interest | 1437 | 51.179 | 9.243 | 33.333 | 45.000 | 51.667 | 58.333 | 66.667 |

**Table 3.** Are the same expressions by different Fed Chairs read differently? This table shows the min-max normalized changes when comparing coefficients using 20-seconds of deepfake video trained on an average of 100,000 iterations of each of the 7 basic emotions. The average emotions displayed through the 20 seconds are derived from 2-second interval screenshots. Using the deepfake technology, the face structure changes but the underlying emotions are the same. We see that fear and surprises displays the lowest changes while disgust shows a much larger variation. Data points highlighted in **bold** show the maximum and minimum differences.

| Emotion | (1) Yellen Original vs Yellen Original Overlaid with Powell's Face | (2) Powell Original vs Powell Original Overlaid with Bernanke's Face | (3) Bernanke Original vs Bernanke Original Overlaid with Yellen's Face |
|---|---|---|---|
| angry | 0.258 | 0.008 | 0.003 |
| disgust | **1.000** | 0.207 | 0.215 |
| fear | **0.000** | 0.190 | 0.008 |
| happy | 0.004 | 0.068 | 0.242 |
| sad | 0.018 | 0.005 | 0.017 |
| surprise | 0.002 | 0.014 | **0.000** |
| neutral | 0.007 | 0.013 | 0.011 |

**Table 4.** Do investors react to negative facial expressions? This table presents coefficients from OLS regressions examining changes in stock (SPY), currency (EUR), (JPY), and the VIX volatility index in response to FOMC chairs' negative emotions and control variables. The analysis includes 1359 to 1404 observations at the minute level spanning 46 FOMC meetings chaired by Ben Bernanke (12), Janet Yellen (16), and Jerome Powell (18) from April 27th, 2011, to September 16th, 2020. Percent changes in SPY, VIX, EUR, JPY are measured over each minute and the absolute value is taken. Negative Facial Expressions represents the intensity of chairs' negative emotions averaged over the preceding three minutes relative to the average across all meetings under the chair. This is to control for the specific nature and disposition of each Fed Chair. Negative Sentiment measures the expressed tone based on FinBERT for each statement. Hawkishness measures the policy stance of chairs based on the keyword list in (Neuhierl and Webet, 2019) and spaCy LLM tokenization. Statement Related measures the frequency of statements in a time interval that are related to the FOMC Press Statement given at 2.00pm. All language parameters are averaged over each rolling minute. Predrift captures percent changes in the 30 minutes from 2.00pm to 2.30pm before the FOMC press conference for SPY, VIX, EUR, JPY respectively. MPU indicates the Monetary Policy Uncertainty index before the FOMC meeting as per Husted et al. (2020). %ΔFDFD denotes the change in Federal Funds Rate on the day of FOMC Press Conference. Standard errors, shown in parentheses, are clustered at the FOMC meeting level. Variable definitions are detailed in Table 1.

|  | (1) %Δ SPY | (2) %Δ SPY | (3) %Δ SPY | (4) %Δ VIX | (5) %Δ VIX | (6) %Δ VIX | (7) %Δ EUR | (8) %Δ EUR | (9) %Δ JPY |
|---|---|---|---|---|---|---|---|---|---|
| Negative Facial | −0.007 | −0.008 | −0.007** | −0.016 | −0.083** | −0.065* | 0.003 | 0.001 | 0.005** |
|  | (0.005) | (0.006) | (0.002) | (0.043) | (0.040) | (0.033) | (0.003) | (0.003) | (0.002) |
| Negative Sentiment | 0.003* | 0.003* | 0.003* | 0.010 | 0.009 | 0.013 | 0.001 | 0.001 | 0.002 |
|  | (0.001) | (0.001) | (0.001) | (0.010) | (0.010) | (0.009) | (0.001) | (0.001) | (0.001) |
| Statement Related | −0.000 | −0.000* | −0.000 | −0.001 | −0.001 | −0.000 | 0.001** | 0.001** | 0.000 |
|  | (0.000) | (0.000) | (0.000) | (0.002) | (0.002) | (0.002) | (0.001) | (0.001) | (0.000) |
| FLS_Ratio | 0.004* | 0.005* | 0.005** | 0.005 | 0.006 | 0.020 | 0.001 | 0.000 | −0.000 |
|  | (0.003) | (0.003) | (0.002) | (0.017) | (0.017) | (0.015) | (0.001) | (0.001) | (0.001) |
| Δ FDFD | −0.011 | −0.008 | 0.000 | 0.088 | 0.040 | 0.000 | −0.012 | 0.000 | 0.007 |
|  | (0.018) | (0.018) | (.) | (0.185) | (0.178) | (.) | (0.015) | (.) | (0.013) |
| MPU | −0.000 |  |  | 0.001 | 0.000 |  | 0.000 | 0.000 | −0.000 |
|  | (0.000) |  |  | (0.000) | (0.000) |  | (0.000) | (.) | (0.000) |
| Predrift SPY | −0.001 | −0.001 | 0.000 |  |  |  |  |  |  |
|  | (0.005) | (0.006) | (.) |  |  |  |  |  |  |
| Hawkish | −0.000 | −0.000 | −0.000 | 0.002 | 0.002 | 0.000 | −0.000*** | −0.000*** | −0.000 |

|  | (0.000) | (0.000) | (0.000) | (0.003) | (0.003) | (0.003) | (0.000) | (0.000) | (0.000) |
|---|---|---|---|---|---|---|---|---|---|
| Public_Interest |  |  | 0.000 |  |  | 0.000 | 0.001*** | 0.000 |  |
|  |  |  | (.) |  |  | (.) | (0.000) | (.) |  |
| Predrift VIX |  |  |  | −0.006 | −0.006 | 0.000 |  |  |  |
|  |  |  |  | (0.007) | (0.007) | (.) |  |  |  |
| Predrift EUR |  |  |  |  |  |  | 0.005 | 0.000 |  |
|  |  |  |  |  |  |  | (0.005) | (.) |  |
| Predrift JPY |  |  |  |  |  |  |  |  | 0.010 |
|  |  |  |  |  |  |  |  |  | (0.011) |
| Chair FE | No | Yes | No | No | Yes | No | No | No | No |
| Meeting FE | No | No | Yes | No | No | Yes | No | Yes | No |
| r^2 | 0.016 | 0.290 | 0.049 | 0.049 | 0.065 | 0.316 | 0.044 | 0.235 | 0.034 |
| N | 1389.000 | 1389.000 | 1359.000 | 1359.000 | 1404.000 | 1359.000 | 1404.000 | 1404.000 | 1404.000 |

Table 5. Do investors react to negative facial expressions? This table presents coefficients from OLS regressions examining changes in stock volume (SPY), currency (EUR) tick count, and the VIX volatility index tick count in response to FOMC chairs' negative emotions and control variables. The analysis includes 1359 to 1404 observations at the minute level spanning 46 FOMC meetings chaired by Ben Bernanke (12), Janet Yellen (16), and Jerome Powell (18) from April 27th, 2011, to September 16th, 2020. Percent changes in SPY, VIX, EUR, JPY are measured over each minute and the absolute value is taken. Negative Facial Expressions represents the intensity of chairs' negative emotions averaged over the preceding three minutes relative to the average across all meetings under the chair. This is to control for the specific nature and disposition of each Fed Chair. Negative Sentiment measures the expressed tone based on FinBERT for each statement . Hawkishness measures the policy stance of chairs based on the keyword list in (Neuhierl and Webet, 2019) and spaCy LLM tokenization. Statement Related measures the frequency of statements in a time interval that are related to the FOMC Press Statement given at 2.00pm. All language parameters are averaged over each rolling minute. Predrift captures percent changes in the 30 minutes from 2.00pm to 2.30pm before the FOMC press conference for SPY, VIX, EUR, JPY respectively. %ΔFDFD denotes the change in Federal Funds Rate on the day of FOMC Press Conference. Standard errors, shown in parentheses, are heteroskedasticity-robust. Variable definitions are detailed in Table 1.

|  | (1) SPY Vol | (2) SPY Vol | (3) VIX Tick | (4) VIX Tick | (5) EUR Tick | (6) EUR Tick | (7) EUR Tick |
|---|---|---|---|---|---|---|---|
| Negative Facial | −0.035 | −0.059*** | −0.453*** | −0.002 | 0.386*** | −0.501*** | −0.068 |
|  | (0.027) | (0.023) | (0.079) | (0.004) | (0.067) | (0.069) | (0.078) |
| Negative Sentiment | 0.019 | 0.022 | 0.004 | −0.003 | 0.168*** | 0.141*** | 0.095*** |
|  | (0.013) | (0.017) | (0.044) | (0.003) | (0.055) | (0.046) | (0.033) |
| Statement Related | −0.004* | −0.003 | −0.005 | 0.001 | 0.028*** | 0.022*** | 0.023*** |
|  | (0.002) | (0.002) | (0.010) | (0.001) | (0.009) | (0.007) | (0.005) |
| Hawkish | −0.002 | −0.001 | 0.002 | 0.000 | −0.000 | −0.003 | −0.007* |
|  | (0.002) | (0.002) | (0.006) | (0.000) | (0.005) | (0.005) | (0.004) |
| Δ FDFD | 0.000 | −0.463*** | −1.475*** | 0.000 | −1.050*** | −1.404*** | 0.000 |
|  | (.) | (0.082) | (0.172) | (.) | (0.213) | (0.180) | (.) |
| Predrift SPY | 0.000 | −0.052 |  |  |  |  |  |
|  | (.) | (0.041) |  |  |  |  |  |
| Predrift VIX |  |  | −0.105*** | 0.000 |  |  |  |
|  |  |  | (0.009) | (.) |  |  |  |
| Predrift EUR |  |  |  |  | 0.697*** | 0.432*** | 0.000 |
|  |  |  |  |  | (0.141) | (0.121) | (.) |
| Chair FE | No | Yes | Yes | No | No | No | Yes |
| Meeting FE | No | No | No | Yes | No | No | Yes |
| $r^2$ | 0.399 | 0.122 | 0.461 | 0.998 | 0.069 | 0.375 | 0.691 |
| N | 1389.000 | 1389.000 | 1359.000 | 1359.000 | 1404.000 | 1404.000 | 1404.000 |

*, **, *** represent significance at the 10%, 5%, and 1% level.

Table 6. Does the Fed Chair consciously control his/her facial expression? This table presents coefficients from OLS regressions examining changes in stock (SPY), currency (EUR), (JPY), and the VIX volatility index in response to FOMC chairs' negative facial expressions and control variables. The analysis includes 1404 observations at the minute level spanning 46 FOMC meetings chaired by Ben Bernanke (12), Janet Yellen (16), and Jerome Powell (18) from April 27th, 2011, to September 16th, 2020. Percent changes in SPY, VIX, EUR are measured over each minute and the absolute value is taken. The dependent variables are the facial expressions. Negative Facial Expressions represents the intensity of chairs' emotions related to the particular emotion averaged over the preceding three minutes relative to the average across all meetings under the chair. This is to control for the specific nature and disposition of each Fed Chair. Negative Sentiment measures the expressed tone based on FinBERT for each statement. Hawkishness measures the policy stance of chairs based on the keyword list in (Neuhierl and Webet, 2019) and spaCy LLM tokenization. Statement Related measures the frequency of statements in a time interval that are related to the FOMC Press Statement given at 2.00pm. All language parameters are averaged over each rolling minute. Predrift captures percent changes in the 30 minutes from 2.00pm to 2.30pm before the FOMC press conference for SPY, VIX, EUR respectively. Standard errors, shown in parentheses, are clustered at the meeting level. Variable definitions are detailed in Table 1.

|  | (1) Negative | (2) Negative | (3) Negative | (4) Neutral | (5) Neutral | (6) Happy | (7) Sad |
|---|---|---|---|---|---|---|---|
| Conference Count | 0.031*** | 0.035*** | 0.000 | −0.059*** | 0.000 | −0.023 | −0.027 |
|  | (0.003) | (0.006) | (.) | (0.013) | (.) | (0.021) | (0.016) |
| Negative Sentiment | −0.011 | −0.016 | −0.008 | 0.030 | 0.039* | 0.037 | −0.024 |
|  | (0.017) | (0.012) | (0.011) | (0.026) | (0.020) | (0.047) | (0.025) |
| Statement Related | −0.001 | −0.003 | −0.000 | 0.004 | 0.003 | −0.010 | 0.006 |
|  | (0.003) | (0.003) | (0.003) | (0.006) | (0.005) | (0.007) | (0.006) |
| $\Delta$ FDFD | −0.162* | −0.199 | 0.000 | −0.297 | 0.000 | 0.175 | 0.818* |
|  | (0.088) | (0.282) | (.) | (0.340) | (.) | (0.835) | (0.441) |
| MPU | −0.001*** | −0.001 | 0.000 | 0.003** | 0.000 | −0.003** | 0.002 |
|  | (0.000) | (0.001) | (.) | (0.001) | (.) | (0.001) | (0.001) |
| Chair FE | No | Yes | No | Yes | No | Yes | Yes |
| Meeting FE | No | No | Yes | No | Yes | No | No |
| r2 | 0.089 | 0.329 | 0.558 | 0.194 | 0.456 | 0.064 | 0.154 |
| N | 1404.000 | 1404.000 | 1404.000 | 1404.000 | 1404.000 | 1404.000 | 1404.000 |

*, **, *** represent significance at the 10%, 5% and 1% level.

**Table 7.** Do investors react to other facial expressions that convey transparency? This table presents coefficients from OLS regressions examining changes in stock (SPY), currency (EUR), (JPY), and the VIX volatility index in response to FOMC chairs' happy facial expressions, transparent facial expressions, sad facial expressions, neutral facial expressions and control variables. The analysis includes 1359 to 1404 observations at the minute level spanning 46 FOMC meetings chaired by Ben Bernanke (12), Janet Yellen (16), and Jerome Powell (18) from April 27th, 2011, to September 16th, 2020. Percent changes in SPY, VIX, EUR are measured over each minute and the absolute value is taken. Transparent Facial Expression represents both neutral and happy. The Facial Expressions represents the intensity of chairs' emotions related to the particular emotion averaged over the preceding three minutes relative to the average across all meetings under the chair. This is to control for the specific nature and disposition of each Fed Chair. Negative Sentiment measures the expressed tone based on FinBERT for each statement. Hawkishness measures the policy stance of chairs based on the keyword list in (Neuhierl and Webet, 2019) and spaCy LLM tokenization. Statement Related measures the frequency of statements in a time interval that are related to the FOMC Press Statement given at 2.00pm. All language parameters are averaged over each rolling minute. Predrift captures percent changes in the 30 minutes from 2.00pm to 2.30pm before the FOMC press conference for SPY, VIX, EUR respectively. %ΔFDFD denotes the change in Federal Funds Rate on the day of FOMC Press Conference. Standard errors, shown in parentheses, are heteroskedasticity-robust. Variable definitions are detailed in Table 1.

|  | (1) %Δ SPY | (2) %Δ SPY | (3) %Δ SPY | (4) %Δ SPY | (5) %Δ VIX | (6) %Δ VIX | (7) %Δ VIX | (8) %Δ EUR | (9) %Δ EUR |
|---|---|---|---|---|---|---|---|---|---|
| Happy Facial | −0.001* |  |  |  |  |  |  |  | −0.002*** |
|  | (0.001) |  |  |  |  |  |  |  | (0.001) |
| Negative Sentiment | 0.003** | 0.003* | 0.003** | 0.003** | 0.011 | 0.008 | 0.013 | 0.001 | 0.001 |
|  | (0.002) | (0.002) | (0.001) | (0.002) | (0.014) | (0.014) | (0.012) | (0.001) | (0.001) |
| Hawkish | −0.000 | −0.000 | −0.000 | −0.000 | 0.002 | 0.001 | 0.000 | −0.000*** | −0.000*** |
|  | (0.000) | (0.000) | (0.000) | (0.000) | (0.003) | (0.003) | (0.003) | (0.000) | (0.000) |
| Statement Related | −0.000** | −0.000** | −0.000 | −0.000** | −0.001 | −0.001 | −0.000 | 0.001** | 0.001** |
|  | (0.000) | (0.000) | (0.000) | (0.000) | (0.002) | (0.002) | (0.002) | (0.001) | (0.001) |
| %Δ FDFD | −0.012* | −0.011 | 0.000 | −0.017*** | 0.059 | 0.026 | 0.000 |  |  |
|  | (0.007) | (0.007) | (.) | (0.007) | (0.066) | (0.065) | (.) |  |  |
| MPUUS_MPU | −0.000*** | −0.000*** | 0.000 | −0.000*** | 0.000 | −0.000 | 0.000 |  |  |
|  | (0.000) | (0.000) | (.) | (0.000) | (0.000) | (0.000) | (.) |  |  |
| Predrift SPY | −0.002 | −0.002 | 0.000 | −0.002 |  |  |  |  |  |
|  | (0.002) | (0.002) | (.) | (0.002) |  |  |  |  |  |
| Neutral Facial |  | 0.003* |  |  |  | 0.012 |  |  |  |
|  | (0.002) |  |  | (0.013) |  |  |  |  |  |
| Transparent Facial |  |  | −0.006*** |  |  |  | −0.068*** |  | −0.002* |
|  |  |  | (0.002) |  |  |  | (0.014) |  | (0.001) |
| Sad Facial |  |  |  | 0.007*** |  |  |  | 0.014 |  |
|  |  |  |  | (0.002) |  |  |  | (0.023) |  |
| Predrift VIX |  |  |  |  | −0.006** | −0.006** |  | 0.000 |  |
|  |  |  |  |  | (0.003) | (0.003) |  | (.) |  |
| Predrift EUR |  |  |  |  |  |  |  | 0.003 | 0.004 |
|  |  |  |  |  |  |  |  | (0.003) | (0.003) |
| Chair FE | Yes | Yes | No | Yes | Yes | Yes | No | Yes | Yes |
| Meeting FE | No | No | Yes | No | No | No | Yes | No | No |
| r2 | 0.031 | 0.032 | 0.285 | 0.042 | 0.042 | 0.042 | 0.315 | 0.315 | 0.064 |
| N | 1389.000 | 1389.000 | 1389.000 | 1389.000 | 1359.000 | 1359.000 | 1359.000 | 1359.000 | 1404.000 |

**Table 8.** Do investors react to facial expressions that contrast with word sentiment? This table presents coefficients from OLS regressions examining changes in stock (SPY), currency (EUR), (JPY), and the VIX volatility index in response to FOMC chairs' negative facial expressions and control variables. The analysis includes 1359 to 1420 observations at the minute level spanning 46 FOMC meetings chaired by Ben Bernanke (12), Janet Yellen (16), and Jerome Powell (18) from April 27th, 2011, to September 16th, 2020. Percent changes in SPY, VIX, EUR are measured over each minute and the absolute value is taken. Negative Facial Expressions represents the intensity of chairs' emotions related to the particular emotion averaged over the preceding three minutes relative to the average across all meetings under the chair. This is to control for the specific nature and disposition of each Fed Chair. Negative Sentiment measures the expressed tone based on FinBERT for each statement. Hawkishness measures the policy stance of chairs based on the keyword list in (Neuhierl and Webet, 2019) and spaCy LLM tokenization. Statement Related measures the frequency of statements in a time interval that are related to the FOMC Press Statement given at 2.00pm. All language parameters are averaged over each rolling minute. Predrift captures percent changes in the 30 minutes from 2.00pm to 2.30pm before the FOMC press conference for SPY, VIX, EUR respectively. Standard errors, shown in parentheses, are clustered at the meeting level. Variable definitions are detailed in Table 1.

|  | (1) %Δ SPY | (2) %Δ SPY | (3) %Δ VIX | (4) %Δ VIX | (5) %Δ EUR |
|---|---|---|---|---|---|
| Negative Facial | −0.007 | −0.006** | −0.085** | −0.054* | −0.003 |
|  | (0.006) | (0.003) | (0.039) | (0.033) | (0.003) |
| Negative Sentiment | 0.004 | 0.002 | 0.005 | 0.014 | −0.001 |
|  | (0.004) | (0.003) | (0.029) | (0.033) | (0.003) |
| Negative Facial * Negative_Sent | −0.001 | 0.000 | 0.003 | −0.002 | 0.002 |
|  | (0.003) | (0.002) | (0.026) | (0.026) | (0.002) |
| Statement Related | −0.000 |  | −0.001 |  | 0.001* |
|  | (0.000) |  | (0.002) |  | (0.001) |
| %Δ FDFD | −0.009 |  | 0.035 |  | −0.017 |
|  | (0.018) |  | (0.184) |  | (0.012) |
| Predrift SPY | −0.001 | 0.000 |  |  |  |
|  | (0.006) | (.) |  |  |  |
| Predrift VIX |  |  | −0.006 | 0.000 |  |
|  |  |  | (0.007) | (.) |  |
| Predrift EUR |  |  |  |  | 0.002 |
|  |  |  |  |  | (0.005) |
| Chair FE | No | Yes | Yes | No | Yes |
| Meeting FE | No | Yes | No | Yes | No |
| r2 | 0.022 | 0.279 | 0.049 | 0.315 | 0.065 |
| N | 1389.000 | 1450.000 | 1359.000 | 1420.000 | 1404.000 |

*, **, *** represent significance at the 10%, 5% and 1% level.

**Table 9.** How do investors interpret negative facial expressions? This table presents coefficients from OLS regressions examining changes in stock (SPY) and the VIX volatility index in response to FOMC chairs' negative facial expressions and control variables. The analysis includes 1404 observations at the minute level spanning 46 FOMC meetings chaired by Ben Bernanke (12), Janet Yellen (16), and Jerome Powell (18) from April 27th, 2011, to September 16th, 2020. Percent changes in SPY, VIX, EUR are measured over each minute and the absolute value is taken. Negative Facial Expressions represents the intensity of chairs' emotions related to the particular emotion averaged over the preceding three minutes relative to the average across all meetings under the chair. This is to control for the specific nature and disposition of each Fed Chair. Negative Sentiment measures the expressed tone based on FinBERT for each statement . Hawkishness measures the policy stance of chairs based on the keyword list in (Neuhierl and Webet, 2019) and spaCy LLM tokenization. Statement Related measures the frequency of statements in a time interval that are related to the FOMC Press Statement given at 2.00pm. All language parameters are averaged over each rolling minute. Cfquart is an indicator, factor variable of which quartile of a Fed Chair's career is that conference in. Congress30 and Congress10 represent whether a FOMC Press Conference was held within 30 days or 10 days after a congressional testimony respective. Predrift captures percent changes in the 30 minutes from 2.00pm to 2.30pm before the FOMC press conference for SPY, VIX respectively. Standard errors, shown in parentheses, are heteroskedasticity-robust. Variable definitions are detailed in Table 1.

| | (1) %Δ SPY | (2) %Δ SPY | (3) %Δ SPY | (4) %Δ SPY | (5) %Δ SPY | (6) %Δ VIX | (7) %Δ VIX | (8) %Δ VIX | (9) %Δ VIX |
|---|---|---|---|---|---|---|---|---|---|
| Negative Facial | −0.005 | −0.000 | −0.004 | −0.009*** | −0.008*** | −0.015 | 0.036 | −0.029 | −0.079*** |
|  | (0.003) | (0.003) | (0.003) | (0.003) | (0.002) | (0.028) | (0.026) | (0.019) | (0.021) |
| Negative Facial i/r Conference Count | −0.000 | −0.001*** | | | | −0.005*** | | | |
|  | (0.000) | (0.000) | | | | (0.002) | | | |
| Negative Sentiment | 0.003* | 0.003* | 0.003* | 0.003* | 0.003* | 0.009 | 0.009 | 0.009 | 0.009 |
|  | (0.002) | (0.002) | (0.002) | (0.002) | (0.002) | (0.014) | (0.014) | (0.014) | (0.014) |
| Statement Related | −0.000** | −0.000** | −0.000** | −0.000** | −0.000** | −0.001 | −0.000 | −0.000 | −0.001 |
|  | (0.000) | (0.000) | (0.000) | (0.000) | (0.000) | (0.002) | (0.002) | (0.002) | (0.002) |
| %Δ FDFD | −0.010 | −0.006 | −0.010 | −0.012* | −0.008 | 0.066 | 0.122* | 0.148* | 0.018 |
|  | (0.008) | (0.008) | (0.008) | (0.007) | (0.007) | (0.068) | (0.070) | (0.078) | (0.067) |
| MPUUS_MPU | −0.000 | | −0.000*** | −0.000*** | −0.000 | 0.000 | 0.001*** | 0.001*** | 0.000 |
|  | (0.000) | | (0.000) | (0.000) | (0.000) | (0.000) | (0.000) | (0.000) | (0.000) |
| Predrift SPY | −0.001 | 0.000 | −0.002 | −0.002 | −0.000 | | | | |
|  | (0.002) | (0.002) | (0.002) | (0.002) | (0.002) | | | | |
| Negative * cfquart | | | | −0.002** | | | | −0.023*** | |
|  |  |  |  | (0.001) |  |  | (0.007) |  |  |
| Negative * congre 30 |  |  |  | 0.001 |  |  |  | 0.038** |  |
|  |  |  |  | (0.002) |  |  |  | (0.019) |  |
| Negative * congre 10 |  |  |  |  | 0.007* |  |  |  | −0.035 |
|  |  |  |  |  | (0.004) |  |  |  | (0.023) |
| Predrift VIX |  |  |  |  |  | −0.006** | −0.005** | −0.007*** | −0.005** |
|  |  |  |  |  |  | (0.003) | (0.002) | (0.002) | (0.003) |
| Chair FE | No | Yes | Yes | Yes | No | Yes | No | No | Yes |
| r2 | 0.012 | 0.028 | 0.039 | 0.037 | 0.014 | 0.054 | 0.017 | 0.013 | 0.050 |
| N | 1389.000 | 1389.000 | 1389.000 | 1389.000 | 1389.000 | 1359.000 | 1359.000 | 1359.000 | 1359.000 |

*, **, *** represent significance at the 10%, 5%, and 1% level.

# Appendix

**A1. FOMC Press Conference and Introductory Statement Recordings List**

| Panel A: List of Publicly Available Recordings of FOMC Press Conference | | |
|---|---|---|
| 4/27/2011 | 6/22/2011 | 11/2/2011 |
| 1/25/2012 | 4/25/2012 | 6/20/2012 |
| 9/13/2012 | 12/12/2012 | 3/20/2013 |
| 6/19/2013 | 9/18/2013 | 12/18/2013 |
| 3/19/2014 | 6/18/2014 | 9/17/2014 |
| 12/17/2014 | 3/18/2015 | 6/17/2015 |
| 9/17/2015 | 12/16/2015 | 3/16/2016 |
| 6/15/2016 | 9/21/2016 | 12/14/2016 |
| 3/15/2017 | 6/14/2017 | 9/20/2017 |
| 12/13/2017 | 3/21/2018 | 6/13/2018 |
| 9/26/2018 | 12/19/2018 | 1/30/2019 |
| 3/20/2019 | 5/1/2019 | 6/19/2019 |
| 7/31/2019 | 9/18/2019 | 10/30/2019 |
| 12/11/2019 | 1/29/2020 | 4/29/2020 |
| 6/10/2020 | 7/29/2020 | 9/16/2020 |
| 11/5/2020 | 12/16/2020 | |

Appendix 1A: List of 47 dates of FOMC Press Conference video recordings publicly available and used in this paper. All press conferences start at 2.30pm EST and we verify the timestamps through the livestream on YouTube. For videos with introductory statements (a feature introduced by Powell), the press conference continues immediately after the statement. I left out March 3rd 2020 due to incompleteness of the data. Press conferences from 29th April 2020 onwards are done via zoom due to Covid restrictions.

**A2. Video Background Analysis**

Due to the large number of screenshots taken at fixed 2-second intervals, I separate out those where reports are asking questions from the Fed Chair using a background analyzer. For instance, on 5th November 2020, screenshot_0827.jpg shows Powell speaking but screenshot_0828 shows the camera panning out to the zoom session. The program correctly identifies using cosine similarity test that screenshot_0828 is not of the Fed Chair, thus, I remove these screenshots from the analysis.

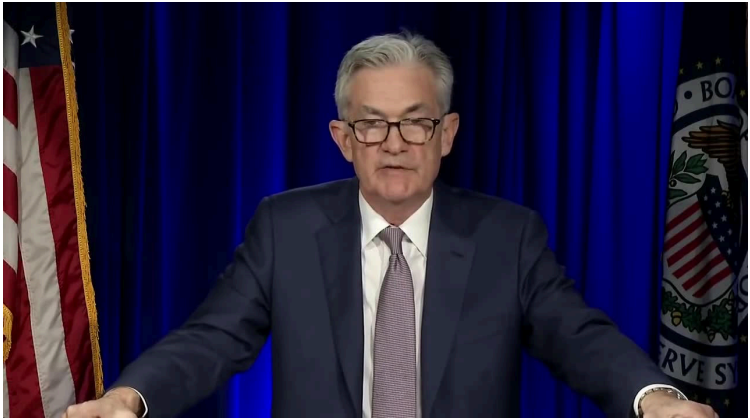

**Panel A.** Fed Chair speaking in *screenshot_0827.jpg*, which the program identifies correctly as an image of Powell speaking, thus, I include it in the analysis.

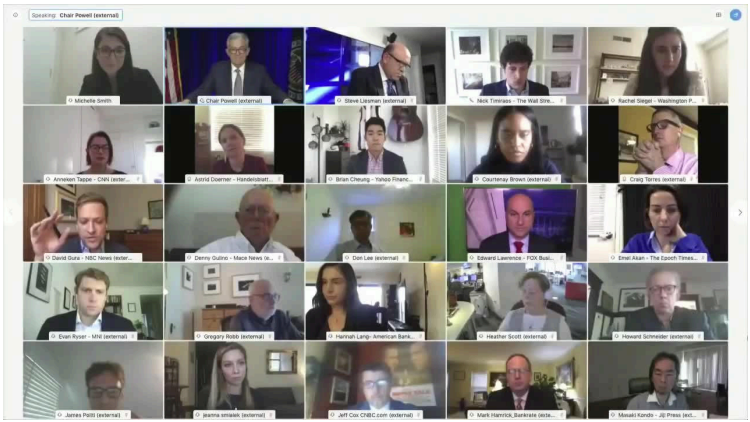

**Panel B.** Reporters asking questions in *screenshot_0827.jpg*, which the program identifies correctly as not an image of Powell speaking, thus, I remove it from the analysis.

**Figure A1.** Comparison of separate screenshots of Fed Chair speaking versus reporters asking questions.

## A3. NLP Keywords

3. To check for hawkish sentiments, I employ the keyword search by Neuhierl and Weber (2019).

| Dovish | Hawkish |
| --- | --- |
| anchor inflation expectations | aggregate demand higher |
| anchored inflation expectations | asset prices increase |
| boost aggregate demand | asset prices rise |
| boost economic activity | business investment increased |
| cut federal funds rate | declines unemployment rate |
| cut interest rates | declining unemployment rate |
| cuts federal funds rate | drop unemployment rate |
| cutting federal funds rate | economic activity increased |
| declines asset prices | economic outlook increased |
| declines crude oil | employment increased |
| declines economic activity | energy prices rise |
| declines employment | exchange rates lower |
| declines energy prices | gradual increases federal funds rate |
| declines house prices | gross domestic product rising |
| declines labor force participation | growing current account deficit |
| declining house prices | higher asset prices |
| declining interest rates | higher employment |
| downward pressure asset prices | higher energy prices |
| downward pressure house prices | higher federal funds rate |
| downward pressure interest rates | higher house prices |
| drop crude oil | higher inflation expectations |
| drop house prices | higher interest rates |
| eased stance monetary policy | higher productivity growth |
| easing monetary policy | higher unit labor costs |
| employment declined | house prices increase |
| employment fallen | house prices increased |
| employment fell | house prices rise |
| employment stable | house prices rising |
| federal funds rate lower | increase asset prices |
| firmly anchored inflation expectations | increase core inflation |
| house prices declined | increase current account surpluses |
| house prices fallen | increase economic activity |
| house prices fell | increase employment |
| increase aggregate demand | increase energy prices |
| increase current account deficit | increase federal funds rate |
| increase labor productivity | increase house prices |
| increase unemployment rate | increase inflation expectations |
| increases productivity growth | increase interest rates |
| increases labor productivity | increase productivity growth |
| increases productivity growth | increase resource utilization |
| inflation expectations anchored | increase target federal funds |
| inflation expectations declined | increase unit labor costs |
| inflation expectations firmly anchored | increased economic activity |
| inflation expectations remained stable | increased employment |
| inflation expectations stable | increased labor force participation |
| inflation expectations well anchored | increases aggregate demand |
| interest rates declined | increases asset prices |
| interest rates drop | increases business investment |
| interest rates easing | increases crude oil |
| interest rates lower | increases employment |
| interest rates lowering | increases energy prices |
| interest rates remain | increases federal funds rate |
| keeping interest rates | increases house prices |
| keeping monetary policy | increases inflation expectations |
| labor productivity increased | increases interest rates |
| lower energy prices | increases output gap |
| lower federal funds rate | increases unit labor costs |
| lower house prices | inflation expectations increased |
| lower inflation expectations | interest rates higher |
| lower interest rates | interest rates increase |
| lower level real oil prices | interest rates increased |
| lower potential output | interest rates might rise |
| lowered federal funds rate | interest rates raise |
| lowering federal funds rate | interest rates raised |
| lowering interest rates | interest rates rise |
| monetary policy easing | interest rates rising |
| nonaccelerating inflation rate | lower current account deficit |
| productivity growth increased | lower productivity growth |
| productivity growth increases | lower unemployment rate |
| raise aggregate demand | monetary policy tightening |
| rapid productivity gains | personal saving rate fallen |
| reduce federal funds rate | raise federal funds rate |
| reduce interest rates | raise interest rates |
| reduce unemployment rate | raised interest rates |
| reduced economic activity | raising asset prices |
| reduced federal funds rate | raising federal funds rate |
| reduced interest rates | raising interest rates |
| reducing federal funds rate | rapid productivity growth |
| reducing interest rates | reduce current account deficit |
| reduction aggregate demand | reductions unemployment rate |
| reduction federal funds rate | resource utilization increased |
| reduction inflation expectations | rise asset prices |
| reduction interest rates | rise core inflation |
| reductions federal funds rate | rise employment |
| reductions interest rates | rise energy prices |
| resource utilization subdued | rise federal funds rate |
| rise productivity growth | rise headline inflation |
| rise unemployment rate | rise house prices |
| rising current account deficit | rise inflation expectations |
| rising productivity growth | rise interest rates |
| risks economic activity | rise personal saving rate |
| risks economic outlook | rise unit labor costs |
| risks outlook economic activity | rising asset prices |
| stabilizing economic activity | rising employment |
| stabilizing employment | rising energy prices |
| stabilizing monetary policy | rising house prices |
| stable economic conditions | rising inflation expectations |
| stable inflation expectations | rising interest rates |
| stable inflation rate | risks long term inflation outlook |
| stable interest rates | sharp increases energy prices |
| stable monetary policy | sharp increases interest rates |
| stable prices moderate | sharp rise interest rates |
| subdued unit labor costs | tightening monetary policy |
| sustainable employment | unemployment rate declining |
| unemployment rate declined | unemployment rate fallen |
| unemployment rate rising | unemployment rate fell |
| upward pressure exchange rates | unemployment rate lower |
| well anchored inflation expectations | upward pressure core inflation |
| | upward pressure interest rates |

2. To check for Statement related sentences, I compile a list of keywords after manually observing a representative sample of the transcripts.

| Statement Related Terms | | |
|---|---|---|
| FOMC statement | earlier statement | stated then |
| todays policy | indicated in the statement | early this afternoon |
| policy statement | released earlier today | conjunction with meeting |
| todays meeting | extensive discussions | what monetary policy might |
| turning to todays meeting | our economic outlook | our symmetric |
| our projection | we expect inflation | our views |
| our operations | our work | our statement |
| our measure | our strategy | contingent on projected |
| growth is expected | committees | committees belief |
| our response | our guess | |

3. To check for forward-looking sentences (FLS), I use FinBERT FLS (Huang, Hui and Yi, 2023).

## A4. Deepfake Video Creation

To create the deepfakes, I use DeepFaceLab, Nvidia RTX3000 series, build 11.20.2021 to train my models. I first identify the source and destination videos from PCA analysis to find the representative videos.

I use a computer specification of GPU - Nvidia GeForce RTX 3060, and a CPU - 12th Gen Intel i7-12700KF, 3610 Mhz. I use the default settings when training the model, except for specific instances where a different setting may outperform the default. For example, I use the *df-ud* model in training the SAEHD framework and use batch size of 4. More details can be found in Figure A2.

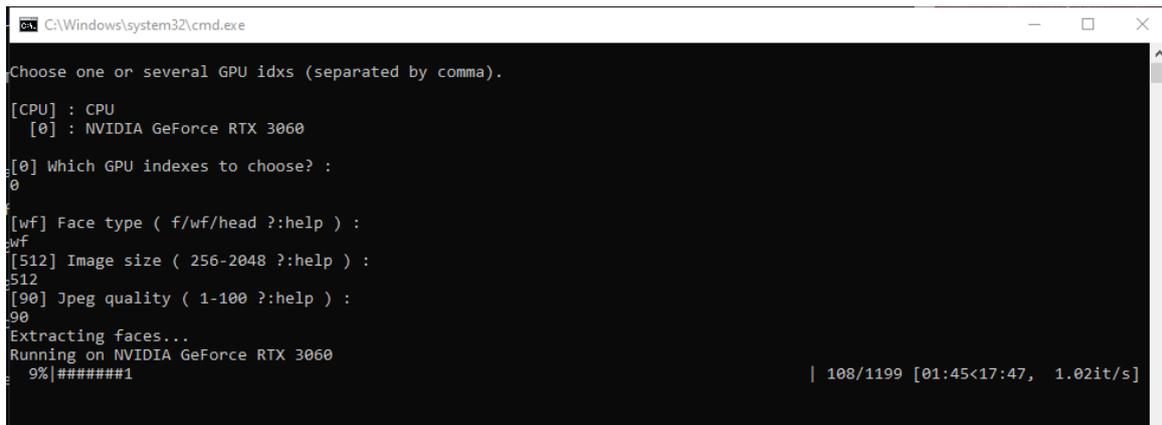

**Panel A:** Specification used to extract the faces from the videos

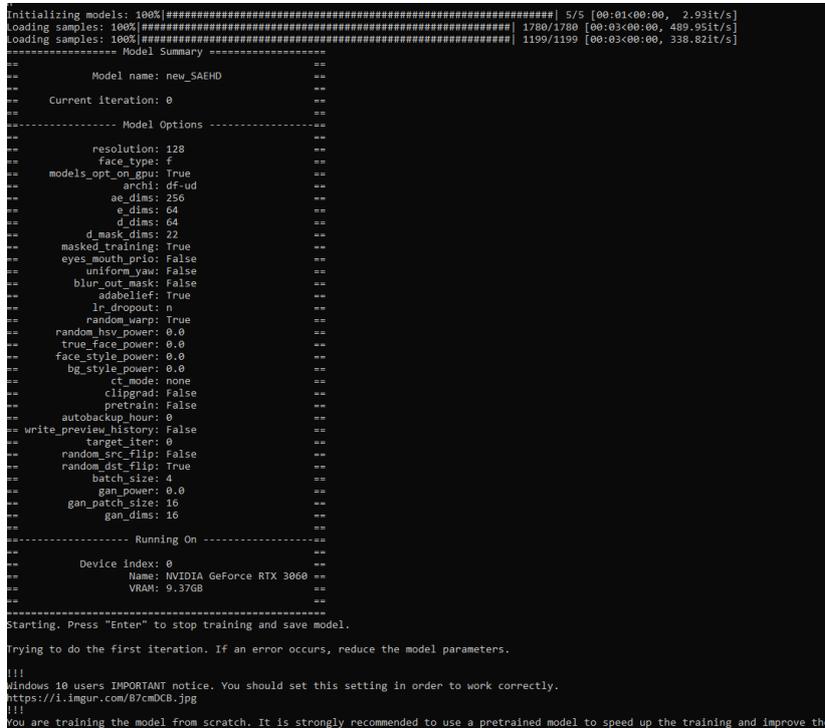

**Panel B.** Comparison of a deepfake of Fed Chair Janet Yellen during FOMC press conference on September 21, 2016 using Fed Chair Powell and their facial analysis result

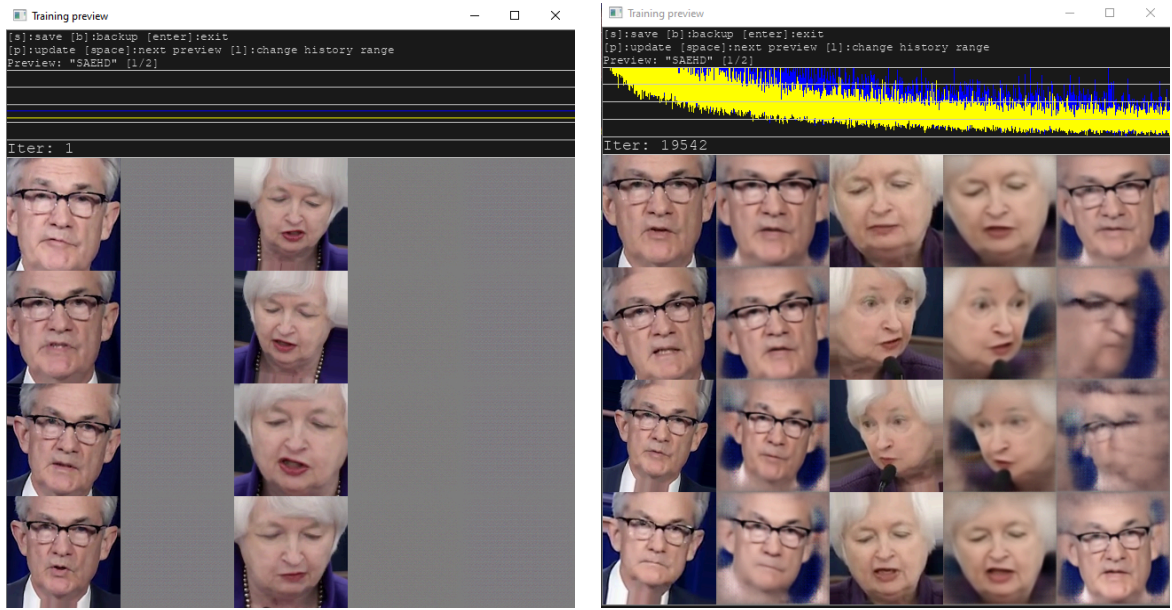

**Panel C.** Training of SAEHD mode, starting from iteration 1. I train to an average of a 100,000 iterations before constructing the deepfake videos.

**Figure A2.** Steps and specifications in training deepfake model.